\documentclass[prl,aps,twocolumn,floats,nofootinbib,showpacs,superscriptaddress]{revtex4}

\usepackage{amsmath,amssymb,dsfont,graphicx,psfrag}
\usepackage{epsfig}
\usepackage{graphicx}

\begin{document}
\title{Superfluid-Insulator transition of quantum Hall domain walls in bilayer graphene}

\author{Victoria Mazo}
\affiliation{Department of Physics, Bar-Ilan University, Ramat-Gan 52900, Israel}

\author{Chia-Wei Huang}
\affiliation{Department of Physics, Bar-Ilan University, Ramat-Gan 52900, Israel}

\author{Efrat Shimshoni}
\affiliation{Department of Physics, Bar-Ilan University, Ramat-Gan 52900, Israel}

\author{Sam T.~Carr}
\affiliation{School of Physical Sciences, University of Kent, Canterbury CT2 7NH, UK}

\author{H.~A.~Fertig}
\affiliation{Department of Physics, Indiana University, Bloomington, IN 47405, USA}

\date{\today}
\begin{abstract}
We consider the zero-filled quantum-Hall ferromagnetic state of bilayer graphene subject to a kink-like perpendicular electric field,
which generates domain walls in the electronic state and low-energy collective modes confined to move along them.
In particular, it is shown that two pairs of collective helical modes are formed at opposite sides of the kink, each pair consisting of modes with identical helicities. We derive an effective field-theoretical model of these modes in terms of two weakly coupled
anisotropic quantum spin-ladders, with parameters tunable through control of the electric and magnetic fields.
This yields a rich phase diagram, where due to the helical nature of the modes, distinct phases possess very different charge conduction properties.
Most notably, this system can potentially exhibit a transition from a superfluid
to an insulating phase.

\end{abstract}
\maketitle

Among the most intriguing electronic properties of graphene is the emergence of novel collective states in the quantum Hall (QH) regime \cite{CastroNetoRMP}.
In particular, the peculiar $\nu=0$ QH states in both monolayer graphene (MLG) \cite{Abanin2007,Checkelsky,Du2009} and bilayer graphene (BLG) \cite{Feldman2009} suggest
that Coulomb interactions lift the degeneracies of the half-filled zero energy Landau level. The multitude of discrete degrees of freedom (two valleys ($\mathbf{K}$,$\mathbf{K'}$) and two spin states in MLG, and an additional layer index in BLG) dictates a rich variety of possible exchange-induced broken symmetry states \cite{Jung2009,Kharitonov_bulk,QHFMGexp}, as generalizations of the spontaneously polarized ferromagnetic state of an ordinary two-dimensional (2D) electron gas \cite{QHFM}. These can be controlled by external fields: in MLG, primarily by tuning the Zeeman energy via a strong parallel magnetic field \cite{Young2013}; in BLG, the orbital isospin degeneracies can be lifted by applying a perpendicular electric field \cite{McCann2006}.

The unique features of the broken symmetry $\nu=0$ QH states are most prominently manifested by the nature of their collective excitations. While in the standard QH ferromagnet the elementary charge excitations are Skyrmions \cite{QHFM}, more complex forms of spin-textures have been predicted in graphene, e.g. charge-$2e$ Skyrmions in BLG \cite{2eSkyrm}. Yet more remarkably, the particle-hole symmetry of the bulk spectrum allows
the formation of charge conducting edge modes associated with
kinks in the effective Zeeman field, where it changes sign across a line.
These can be realized near physical edges of the graphene ribbon \cite{BreyFertig,Abanin_2006,Mazo},
or in the interior of a BLG sheet subject to non-uniform gating \cite{Martin2008,Paramekanti,Huang}.

The coherent domain wall (DW) forming in the spin/isospin configuration near
such a kink supports a gapless collective mode, which possesses a one-dimensional (1D) dynamics along the kink
of a {\it helical} character. The latter arises from the constraint relating a spin/isospin texture
to the charge degree of freedom \cite{QHFM}. This
yields a mapping to a helical Luttinger liquid (HLL), with a single flavor encoding spin and charge related by duality, in analogy with the edge states of 2D topological insulators (TI) \cite{TIreview,HLL}. However in distinction from the latter, DW modes are not topologically protected by time-reversal symmetry and are therefore not immune to backscattering due to perturbations which violate spin or isospin conservation. Their conduction properties depend crucially on the Luttinger parameter, which is sensitive to the ratio between the effective Zeeman energy and exchange interaction, and may be tuned by external fields \cite{BreyFertig,SFP}. Interestingly, this may trigger a transition from an insulator to a conducting phase manifesting the quantum spin Hall effect (QSHE) \cite{Kharitonov_edge,Young2013}.

\begin{figure}[h]
\begin{center}
\includegraphics[width=0.7\linewidth]{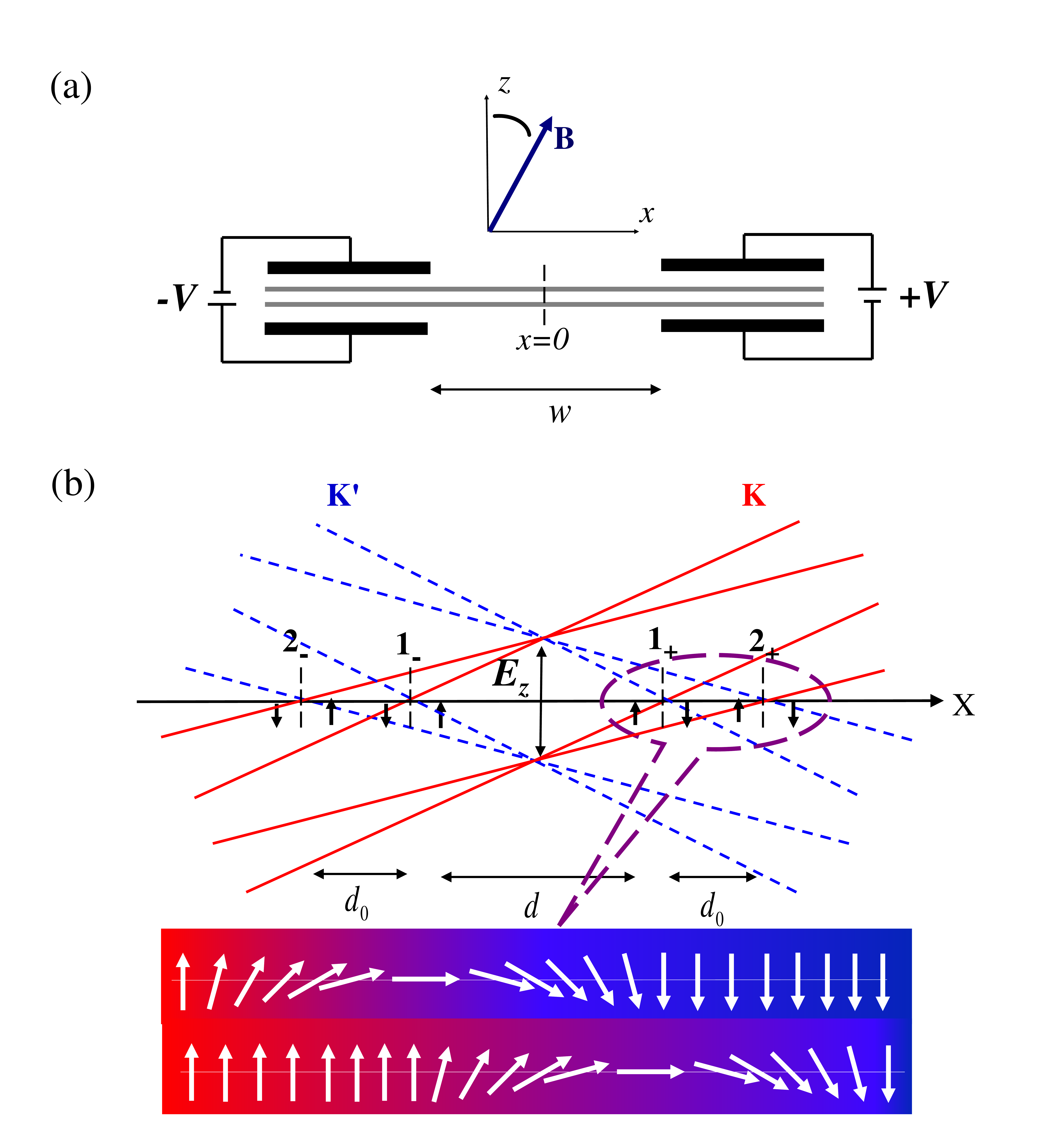}
\caption{(color online) (a) BLG in a split-double-gated setup; the $y$-axis is perpendicular to the page.
(b) The non-interacting energy levels crossing at $\epsilon=0$ vs. guiding center $X$. Arrows denote spin ($S^z$ along ${\bf B}$), full red lines correspond to valley $\mathbf{K}$ and dashed blue lines to valley $\mathbf{K'}$. The blow-up presents a typical $S^z$-configuration of a coupled DW-pair.}\label{dgated_BLG}
\end{center}
\end{figure}

In this paper, we propose a realization of QH DW states in BLG subject to a kink in the perpendicular electric field (see Fig. 1), where pairs of HLL with identical helicities couple to form strongly correlated 1D collective modes. Depending on the external fields manipulating their Luttinger parameter $K$ and coupling strengths, these correlated modes exhibit a charge-density wave (CDW) or superfluid (SF) character, marked by distinct transport properties. We suggest a measurement of the ``antisymmetric conductance" as a particularly revealing probe of transitions between the two phases.

We consider a split-double-gated geometry as discussed in \cite{Martin2008,Paramekanti,Huang}, where the inter-layer bias $V(x)$ imposed on a BLG changes from $+V$ to $-V$ over a distance $w$ along the $x$
direction [Fig. \ref{dgated_BLG}(a)]. In addition, a strong tilted magnetic field ${\bf B}$ enforces Landau quantization, as well as a Zeeman energy $E_z\sim |{\bf B}|$. The single-particle zeroth Landau levels separate into eight different levels
which cross at four distinct guiding center coordinates $X_{n_\pm}$ ($n=1,2$), symmetric
about the point $V(x)=0$ [see Fig. \ref{dgated_BLG}(b)]. This results in
two pairs of parallel 1D channels propagating along the $y$-axis, marked by distinct helicities:
$h={\rm sgn}(X_{n_h})$.

Exchange interactions modify this picture, forming a spin-valley DW structure as depicted in Fig. \ref{dgated_BLG}(b) and replacing the single-particle helical states by mutually coupled collective modes. As explained in earlier work \cite{BreyFertig,SFP,Mazo2}, the quantum dynamics of
each DW-mode (along the $y$-direction) can be described in terms of an effectively 1D spin-$1/2$ field (${\bf S}_{n_h,y}$),
encoding both spin and charge degree of freedom: $S^x$ and $S^y$ are associated with electric charge (a property which descends from the
spin-charge coupling inherent in quantum Hall ferromagnets \cite{QHFM}), and the $S^z$ operator coincides with the electric {\it current}.
Note that the latter correspondence depends on helicity, $S^z_{n_h}\sim hj_e$.
The coupling between adjacent DWs depends on their spacing [$d,d_0$ in Fig. \ref{dgated_BLG}(b)]: $d\approx w\frac{E_z}{2eV}$, $d_0\approx d\frac{\omega_c^2}{\gamma_1^2-\omega_c^2}$, where $\gamma_1$ is the interlayer hopping and $\omega_c=\sqrt{2}\hbar v_F/\ell$ (with $\ell=\sqrt{c\hbar /eB_\perp}$ the magnetic length) \cite{Huang}, so that typically $d_0\ll d$. Hence modes with the same helicity are more strongly coupled. The system is then modeled by the effective Hamiltonian
\begin{equation}\label{2ladders}
H=\sum_{h=\pm}H_h+H_{+-}\; ,
\end{equation}
where the weak coupling $H_{+-}$ will be specified later on, and $H_h$ describe anisotropic spin-$1/2$ two-leg ladders:
\begin{eqnarray}\label{spinladder}
H_h&=&\sum_{n=1,2}H_{n_h}+H_{\perp}^{(h)}\; , \\ \nonumber
H_{n_h}&=&\sum\limits_{y}\left[\frac{J_n^{xy}}{2}\big(S^+_{n_h,y}S^-_{n_h,y+1}+h.c. \big)
+J_n^zS^z_{n_h,y}S^z_{n_h,y+1}\right]\\ \nonumber
H_{\perp}^{(h)}&=&\sum\limits_{y}\left[\frac{J_{\perp}^{xy}}{2}\big(S^+_{1_h,y}S^-_{2_h,y}+h.c. \big)
+J_{\perp}^zS^z_{1_h,y}S^z_{2_h,y}\right]\; .
\end{eqnarray}
The dependence of $J_n^\alpha,J_{\perp}^\alpha$ on the original system parameters and external fields is complicated; however, the anisotropy factors $\Delta_{n(\perp)}\equiv \frac{J_{n(\perp)}^z}{J_{n(\perp)}^{xy}}$ qualitatively reflect the ratio of kinetic energy ($\propto V$) to exchange interaction ($\sim e^2/\ell$). Note that the {\it signs} of $J_{\perp}^\alpha$ depend on the spatial overlap of DWs, and are hereon assumed arbitrary.

We next employ standard Bosonization to express the spin operators in terms of Bosonic fields $\phi_{n_h}(y)$ and their dual $\theta_{n_h}(y)$ \cite{giamarchi}:
$S^+\sim e^{-i\theta}[(-)^y+\cos(2\phi)]$, $S^z\sim [-\partial_y\phi+(-)^y\Lambda\cos(2\phi)]$
with $\Lambda\sim 1/\ell$ a short distance cutoff.
Defining symmetric and antisymmetric modes in each ladder [$\phi_{s_h}=(\phi_{1_h}+\phi_{2_h})/2$, $\theta_{s_h}=(\theta_{1_h}+\theta_{2_h})$ and $\phi_{a_h}=(\phi_{1_h}-\phi_{2_h})$, $\theta_{a_h}=(\theta_{1_h}-\theta_{2_h})/2$, respectively], the leading
continuum limit of $H_h$ becomes ($\hbar=1$)
\begin{equation}\label{Hh}
H_h=\sum_{\nu=s,a}[H_0^{(\nu_h)}+H_{int}^{(\nu_h)}]+H_{as}^{(h)}
\end{equation}
where under the assumption $J_{\perp}^\alpha,|J_1^\alpha-J_2^\alpha|\ll J_n^\alpha$,
\begin{eqnarray}\label{HhBos}
H_0^{(\nu_h)}&=&\frac{v_\nu}{2\pi}\int dy\Big[K_{\nu}(\partial_y\theta_{\nu_h})^2
+\frac{1}{K_{\nu}}(\partial_y\phi_{\nu_h})^2\Big] \\ \nonumber
H_{int}^{(s_h)}&=&{\tilde g}_z\Lambda^2\int dy\,\cos(4\phi_{s_h})\\ \nonumber
H_{int}^{(a_h)}&=&g_{xy}\Lambda^2\int dy\,\cos(2\theta_{a_h})+g_z\Lambda^2\int dy\,\cos(2\phi_{a_h})
\\ \nonumber
H_{as}^{(h)}&=&\int\frac{dy}{2\pi} \big[g_{as}^{xy}(\partial_y\theta_{a_h})(\partial_y\theta_{s_h})
+g_{as}^{z}(\partial_y\phi_{a_h})(\partial_y\phi_{s_h})\big]\; ;
\end{eqnarray}
here $v_{\nu}\sim J_{xy}/\Lambda$ with $J_{\alpha}\equiv\frac{J_1^{\alpha}+J_2^{\alpha}}{2}$, $g_\alpha,{\tilde g}_\alpha\sim J_{\perp}^\alpha /\Lambda$,
\begin{equation}\label{vKdef}
K_s \approx \frac{K}{2}\left(1-\frac{KJ_{\perp}^z}{2\pi v\Lambda}\right)\; ,\quad K_a\approx 2K\left(1+\frac{KJ_{\perp}^z}{2\pi v\Lambda}\right)
\end{equation}
where $K=\frac{\pi}{2\arccos\Delta}$, $\Delta\equiv \frac{-J_z}{J_{xy}}$. Since $g_{as}^\alpha\sim (J_1^\alpha-J_2^\alpha)/\Lambda\ll v$, the marginal last term in Eqs. (\ref{Hh}),(\ref{HhBos}) can be treated perturbatively \cite{as_corr} and is henceforth neglected.

Under the above approximation, the $s$ and $a$ modes are decoupled: $H_h=H_{s}^{(h)}+H_{a}^{(h)}$, with $H_\nu^{(h)}=H_0^{(\nu_h)}+H_{int}^{(\nu_h)}$,
 as in a standard spin-$1/2$ ladder \cite{Shelton,GNTbook}. However, due to the helical nature of the channels $H_{s}^{(h)}$,$H_{a}^{(h)}$ describe the conduction properties of the charge degree of freedom as well: $\partial_y\phi_{\nu_h}$ denote spin-density fluctuations, and also encode the total and relative electric current operators through channels $1_h,2_h$:
\begin{equation}\label{JaJs}
J_{1_h}+J_{2_h}=\frac{-2evh}{\pi K}\partial_y\phi_{s_h}\,,\; J_{1_h}-J_{2_h}=\frac{-evh}{\pi K}\partial_y\phi_{a_h}\,.
\end{equation}
The dual fields $\partial_y\theta_{\nu_h}$ encode the corresponding charge density operators.

The behavior of the symmetric mode is controlled by a sine-Gordon model
$H_{s}^{(h)}$, where the cosine term
is irrelevant for $K_s>1/2$ [e.g., for $K=1$ and arbitrary $J_{\perp}^z<0$; see Eq. (\ref{vKdef})], in which case it is simply a Luttinger liquid \cite{giamarchi,GNTbook}. In contrast, $H_{a}^{(h)}$ contains two competing cosine terms; for arbitrary $K_a$, at least one of them is relevant. In particular, both terms are relevant for $1/2<K_a<2$. This regime includes the self-dual point $K_a=1$, where the model was shown to exhibit an Ising ($Z_2$) quantum phase transition \cite{Shelton,GNTbook,SDSG} from a phase with ordered $\phi_{a_h}$ to an ordered $\theta_{a_h}$. Below we argue that this behavior persists throughout the entire range $1/2<K_a<2$, and discuss the interpretation of the two phases as SF and CDW, respectively \cite{Atzmon}.

For $K_a=1$, the model $H_{a}^{(h)}$ can be exactly mapped to massive free Fermions
\cite{Shelton,GNTbook}. In terms of two species of right and left moving
Majorana fields $\xi_R^\pm$, $\xi_L^\pm$, one obtains
\begin{equation}\label{free_Majo}
H_{a}^{(h)}=\sum_{\tau=\pm}\int dy\left\{\frac{iv_a}{2}(\xi_L^\tau\partial_y\xi_L^\tau-\xi_R^\tau\partial_y\xi_R^\tau)
-im_\tau\xi_R^\tau\xi_L^\tau\right\}\,,
\end{equation}
$m_\pm =\Lambda(g_{xy}\pm g_z)$. This represents two independent Ising chains ($\tau=\pm$) in a transverse field, which possess quantum critical points at $g_{xy}=\pm g_z$ (depending on the relative sign of $g_{xy}$, $g_z$) where one of the masses $m_\tau$ vanishes. The phases separated by this critical point are related by duality: for $|g_{xy}|> |g_z|$, the original field $\theta_{a_h}$ acquires a fixed value ($\pi$ or $0$ for $g_{xy}>0$, $g_{xy}<0$ respectively) and $\phi_{a_h}$ is disordered; for $|g_{xy}|<|g_z|$, the roles of $\theta_{a_h}$, $\phi_{a_h}$ and $g_{xy}$, $g_z$ are interchanged.

\begin{figure}[h]
\begin{center}
\includegraphics[width=0.7\linewidth]{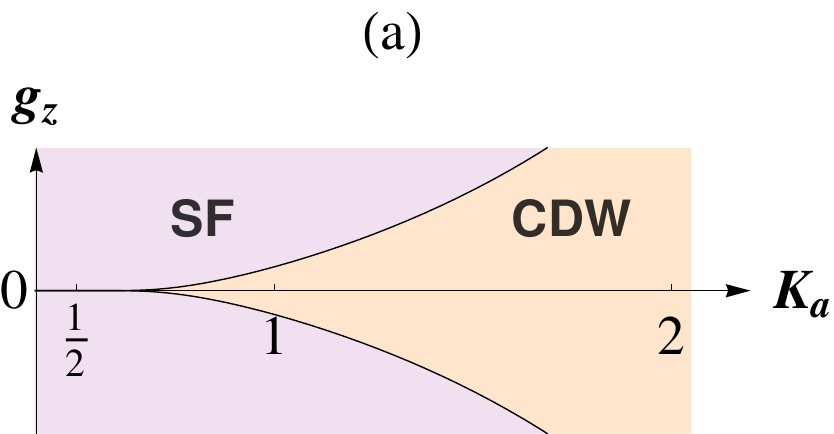}
\includegraphics[width=0.7\linewidth]{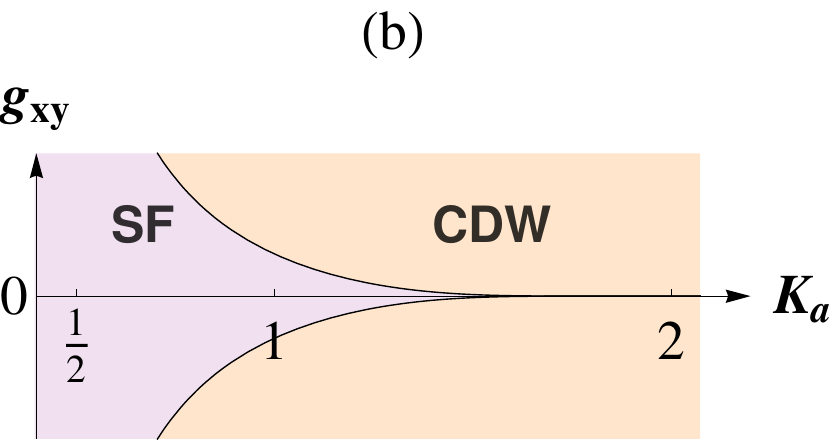}
\caption{Phase diagram of the model $H_{a}^{(h)}$; the phase boundary is derived from Eq. (\ref{QCPvsK}), (a) for $g_{xy}=0.1$,
(b) for $g_{z}=0.1$.}\label{Phase_diagram}
\end{center}
\end{figure}

We next consider a deviation from the self-dual point, $K_a=1+g$ ($-\frac{1}{2}<g<1$). The free model Eq. (\ref{free_Majo}) acquires an interaction term which couples $\tau=\pm$:
\begin{equation}\label{H_g}
H_g=-g\Lambda^2\int dy\xi_R^+\xi_L^+\xi_R^-\xi_L^-\; .
\end{equation}
However, one of the sectors is always more massive, and does not undergo a transition. This justifies a mean-field approximation for the other sector (denoted $\tau_c$), where the operator $\xi_R^{-\tau_c}\xi_L^{-\tau_c}$ is replaced by its expectation value \cite{GNTbook}. The resulting approximation for $H_g$ merely yields a shift of $m_{\tau_c}$ by $\delta m=ig\Lambda^2\langle\xi_R^{-\tau_c}\xi_L^{-\tau_c}\rangle$. Consequently, the critical point determined by $m_{\tau_c}=0$ is shifted but maintains its $Z_2$ character. This yields a transition {\it line}, which can be derived by equating the effective masses of the two competing order fields. The scaling of these masses with $K_a$ \cite{giamarchi} implies that this occurs at
\begin{equation}\label{QCPvsK}
|g_{xy}|^{\frac{1}{2-1/K_a}}\sim |g_z|^{\frac{1}{2-K_a}}\; .
\end{equation}
The resulting phase diagram is depicted in Fig. \ref{Phase_diagram}. The CDW phase is characterized by a gap $\Delta_c\sim m_{\tau_c}$ to fluctuations in the relative charge fields $\theta_{a_h}$, while the SF phase exhibits a gap $\Delta_s$ to fluctuations in $\phi_{a_h}$.

To derive the conduction properties characterizing the distinct phases, we first introduce local coupling terms between the channels $1_h,2_h$ which break translation invariance in the $y$-direction, and are necessary to induce non-trivial transport coefficients. As a minimal choice of such terms \cite{general_disorder}, we consider defects at $y=0$ which add a local correction $J_0$ to $J^{xy}_\perp$ [Eq. (\ref{spinladder})], and a spin-flip term allowing backscattering between the closest channels of opposite helicities [$H_{+-}$ in Eq. (\ref{2ladders})]:
\begin{equation}\label{H_local}
\delta H^{(h)}=J_0\big[S^+_{1_h,0}S^-_{2_h,0}+h.c. \big]\; ,\quad H_{+-}=J\big[S^+_{1_-,0}S^-_{1_+,0}+h.c. \big]\; .
\end{equation}
In terms of Bosonic fields, these yield
\begin{eqnarray}\label{H_local_Bos}
\delta H &=&\sum_{h=\pm}J_0\Lambda\cos[\theta_{1_h}(0)-\theta_{2_h}(0)] \\ \nonumber
&+&\sum_{n,n^\prime=1,2}J_{n,n^\prime}\Lambda\cos[\theta_{n_+}(0)-\theta_{n^\prime_-}(0)]
\end{eqnarray}
where $J_{n,n^\prime}$ with $n,n^\prime=2$ are generated to second order in the perturbations Eq. (\ref{H_local}).

\begin{figure}[h]
\begin{center}
\includegraphics[width=0.7\linewidth]{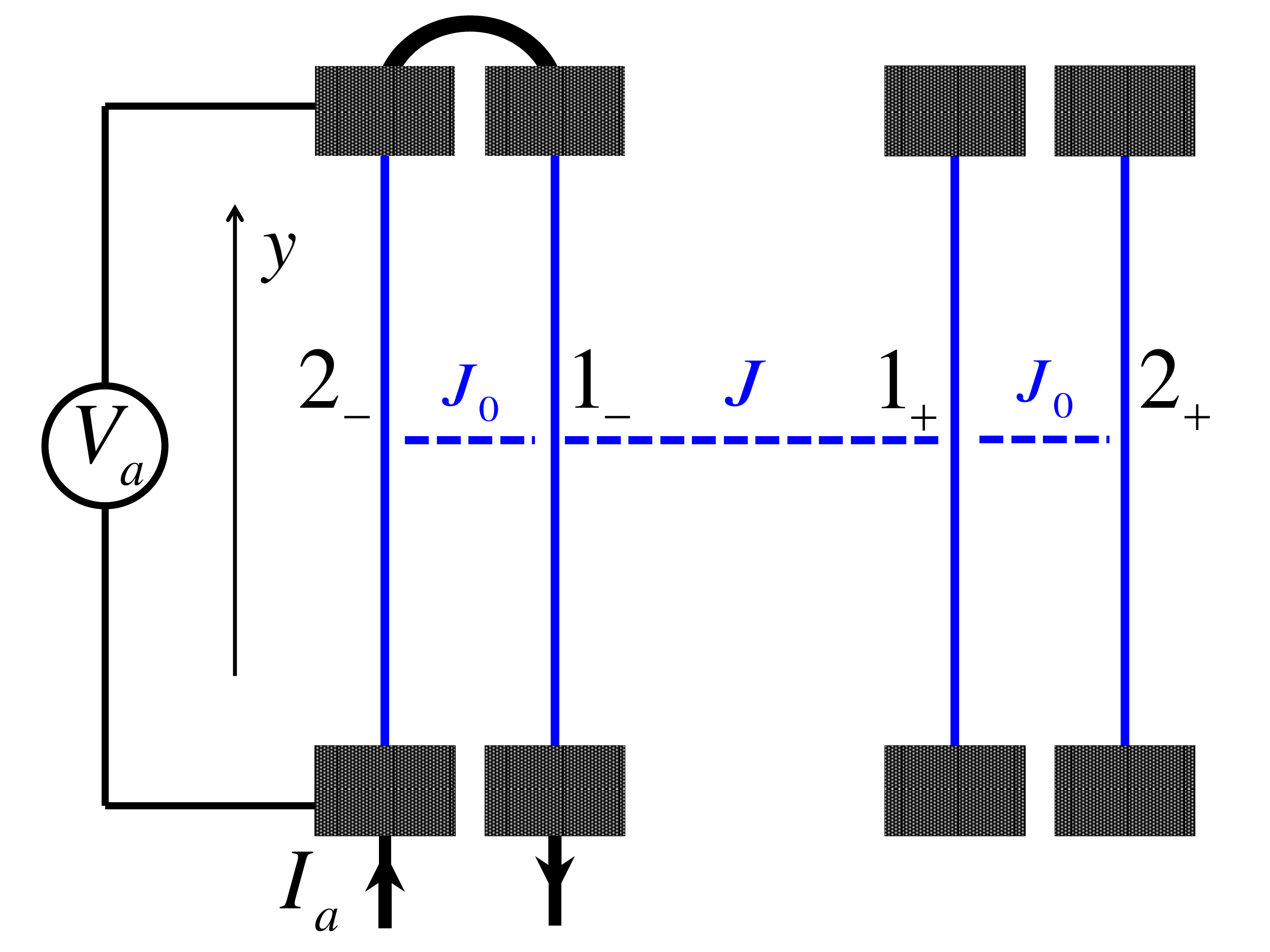}
\caption{Schematic transport measurement geometry, illustrated for
antisymmetric conductance.  Blue solid lines represent DW's. $J$, $J_0$ are defined in Eq. (\ref{H_local}).}\label{G_a}
\end{center}
\end{figure}

We now consider a multi-terminal contact to an external circuitry where current can be driven along the $y$-direction of the BLG sample (Fig. \ref{G_a}). We particularly focus on two observables:
the total two-terminal conductance $G$, and the ``antisymmetric conductance" $G_a=I_a/V_a$ where $I_a$ is a counter-propagating current in the channels $1_h,2_h$, short-circuited at one edge (see Fig. \ref{G_a}). From Kubo's formula,
$G$ is given by the retarded correlation function of the fully symmetric current $J_s=\sum_{h=\pm,n=1,2}J_{n_h}$; similarly, $G_a$ is dictated by the correlation of relative current operators [second term of Eq. (\ref{JaJs})].

We first consider the behavior of $G(T)$ at a finite temperature $T$. The main contribution to the scattering of the current $J_s$ arises from the second term in Eq. (\ref{H_local_Bos}), which couples the $a$ and $s$ modes via the operators
\begin{equation}\label{H+_as}
{\cal O}_\pm=\cos\left[\frac{(\theta_{s_+}(0)-\theta_{s_-}(0))}{2}\right]
\cos\left[\theta_{a_+}(0)\pm\theta_{a_-}(0)\right]\; .
\end{equation}
To leading order in $\delta H$, the conductance $G$ (in units of $e^2/2\pi\hbar$) is then given by \cite{Chap7}
\begin{eqnarray}\label{G_Kubo}
G &=& 4-\delta G\; ,\quad \delta G \sim \int_0^\infty dt\,t\langle[F_\pm(t),F_\pm(0)]\rangle \nonumber \\
& &{\rm where}\quad F_\pm \equiv i[J_s,{\cal O}_\pm]\; .
\end{eqnarray}

In the CDW phase, since $\theta_{a_h}$ are ordered, the second cosine in Eq. (\ref{H+_as}) can be replaced by its finite expectation value and
${\cal O}_\pm\sim\cos\theta$, $\theta\equiv\frac{(\theta_{s_+}(0)-\theta_{s_-}(0))}{2}$. Eq. (\ref{G_Kubo}) with $J_s$ related to $\phi_{s_h}$ via Eq. (\ref{JaJs}) hence yields \cite{KaneFisher,giamarchi}
\begin{equation}\label{G_I_highT}
\delta G\sim T^{\frac{1}{4K_s}-2}\; .
\end{equation}
For accessible values of $K_s$, this typically diverges at low $T$ implying a breakdown of the weak backscattering approximation. The system therefore exhibits an insulating behavior, $G(T\rightarrow 0)=0$. The finite low $T$ dependence of $G$ can be evaluated perturbatively in the dual tunneling operator $\cos(4\phi)$ \cite{KaneFisher,Saleur,Sam_long}, resulting in
\begin{equation}\label{G_I_lowT}
G\sim T^{16K_s-2}\; .
\end{equation}

In the SF phase, $\theta_{a_h}$ are disordered and the correlations of $e^{\pm i\theta_{a_h}}$ yield an exponential decay of $\delta G$ evaluated from Eq. (\ref{G_Kubo}) for $T\ll \Delta_s$. The leading backscattering is therefore governed by {\it second} order terms generated by $\delta H$, which decouple the $a$-mode \cite{OG,Atzmon,Sam_long}, of the form $\cos(2\theta)$. One obtains $\delta G\sim T^{\frac{1}{K_s}-2}$, which under our assumption $K_s>1/2$ still implies an {\it insulating} behavior at $T\rightarrow 0$; the same procedure leading to Eq. (\ref{G_I_lowT}) yields
\begin{equation}\label{G_S_lowT}
G\sim T^{4K_s-2}
\end{equation}
(which approaches a non-universal constant for $K_s\sim 1/2$). We therefore conclude that a transition from SF to CDW is manifested in $G(T)$ as a jump in the power-law $G\sim T^\kappa$, from $\kappa=4K_s-2$ to $\kappa=16K_s-2$.

A more dramatic signature of the SF/CDW transition is expected in the $T$-dependence of $G_a$, which probes the response to a pure antisymmetric current
$I_a$. Backscattering in this channel is solely due to the first term in Eq. (\ref{H_local_Bos}), which can be cast as
\begin{equation}\label{O_a}
{\cal O}_a=J_0\Lambda\sum_{h=\pm}
\cos\left[2\theta_{a_h}(0)\right]\; .
\end{equation}
In the SF phase, we evaluate the deviation $\delta G_a$ from perfect conductance ($G_a=1-\delta G_a$ for each ladder $h=\pm$) from Eq. (\ref{G_Kubo}) with ${\cal O}_\pm$ replaced by ${\cal O}_a$, associated with the disordered operators in this phase. For $T\ll \Delta_s$,
\begin{equation}\label{G_a_S}
\delta G_a\sim \exp\left(-\frac{\Delta_s}{T}\right)
\end{equation}
which implies an exponentially small voltage drop $V_a\sim \delta G_a$ in the setup depicted in Fig. \ref{G_a}. A similar calculation, with ${\cal O}_a$ replaced by its dual $\sum_h\cos\left[2\phi_{a_h}(0)\right]$, yields an exponentially small {\it conductance} in the CDW phase:
\begin{equation}\label{G_a_I}
G_a\sim \exp\left(-\frac{\Delta_c}{T}\right)\; ,
\end{equation}
with $\Delta_c$ a charge gap characterizing this phase. We thus predict that $G_a$ would exhibit a true ``superconductor-insulator" transition, indicated by a jump of $G(T\rightarrow 0)$ from 1 to 0 upon tuning of, e.g., $K_a$ (which monotonically increases with the physical parameter $e/\ell V\sim \sqrt{B_\perp}/V$) through the phase boundaries of Fig. \ref{Phase_diagram}.

In summary, we have shown that pairs of DW's forming in the $\nu=0$ QH state of BLG subject to
a kink-like gate potential $V(x)$ provide a unique realization of spin-$1/2$ ladders, where linkage between the spin and charge degrees of freedom implies that distinct phases of the spin system are distinguishable by electric transport properties. In particular, we propose that in a sufficiently strong magnetic field ${\bf B}$ where $E_z$ is appreciable \cite{Young2013}, the tuning of $V$ or the size and tilt-angle of ${\bf B}$ can induce a SF-CDW transition, clearly observable in the low-$T$ conductance.

As a final remark, note that the paired DW's discussed in our case are apparently analogous to a ``helical ladder" formed by coupling two parallel edge states of TI, with a crucial distinction: in the latter, a coupling in the form of the first term in $H_{\perp}^{(h)}$ [Eq. (\ref{spinladder})] is forbidden. The second term of $H_{\perp}^{(h)}$, analogous to a Josephson coupling resulting from electron-pair tunneling between the HLL's, likewise competes with a term generating a CDW order. However, the latter has a different scaling dimension. Hence, the resulting SF-CDW transition is of a different nature.  This will be discussed in more detail elsewhere \cite{future}.

We thank E. Berg, T. Pereg-Barnea and A. Young for useful discussions. E. S. is grateful to the hospitality of the Aspen Center for Physics (NSF Grant No. 1066293) and to the Simons Foundation. This work was supported by the US-Israel Binational Science Foundation (BSF) grant 2008256,
the Israel Science Foundation (ISF) grant 599/10, and by NSF Grant No. DMR-1005035.


\begin{thebibliography}{235}
\bibitem{general_disorder}
Our analysis can be straightforwardly modified to acount for a more general type of disorder, e.g. a finite density of random impurities \cite{giamarchi}, yielding qualitatively similar results.
\end{thebibliography}
\end{document}